\documentstyle[epsfig,11pt]{article}

\def\be{\begin{equation}}
\def\ee{\end{equation}}
\def\bea{\begin{eqnarray}}
\def\eea{\end{eqnarray}}

\begin{document}

\vspace*{-1.8cm}
\begin{flushright}
{\bf LAL 04-100}\\
\vspace*{0.1cm}
{October 2004}
\end{flushright}
\vspace*{0.8cm}

\begin{center}
{\Large\bf EXPERIMENTAL IMPLICATIONS FOR A LINEAR\\
\vspace*{0,3cm}
 COLLIDER OF SUSY DARK MATTER 
 SCENARIO}
\vspace*{1.cm} 

{\Large{\bf Zhiqing ZHANG }}

\vspace*{0.8cm}
{\large\bf Laboratoire de l'Acc\'el\'erateur Lin\'eaire}\\

 IN2P3-CNRS et Universit\'e de Paris-Sud, B\^at. 200, BP 34, F-91898 Orsay Cedex
\end{center}
\vspace*{0.6cm}


\begin{abstract}
 This talk presents the detection issues for the lightest slepton 
 $\tilde{\tau}_1$ at a future $e^+e^-$ TeV collider given the dark 
 matter constraints set on the SUSY mass spectrum by the WMAP results. 
 Two methods for measuring its mass $m_{\tilde{\tau}_1}$ and the resulting
 precision on the dark matter density are briefly discussed
 in the SUSY mSUGRA scenario with $R$-parity conservation when
 the mass difference between $m_{\tilde{\tau}_1}$
 and that of the lightest neutralino is small (a few GeV).
 The analysis is performed with TESLA parameters in both head-on and
 crossing angle modes.
\end{abstract}
  
\vspace*{0.5cm}

The talk is based on recent studies [1,2] motivated 
by the increasing awareness in the community of the role of an $e^+e^-$ 
Linear Collider (LC) for a precise determination of the SUSY parameters 
which are needed to interpret the dark matter (DM) content of the universe. 
After the WMAP results [3] leading to an accuracy at the $10\%$ level
or $0.094<\Omega_{\rm DM} h^2<0.129$ in two standard deviation range and 
awaiting for the Planck mission in $2007$ which aims at $2\%$, 
it seems appropriate to check that a LC can do its job properly on 
this essential topic. 

In the SUSY scenario with $R$-parity conservation, the lightest SUSY particle 
(LSP) is the lightest neutralino $\chi$. This particle is considered as 
the best candidate to satisfy the cosmological constraints on DM in 
the universe. DM constraints have been recently re-examined [4]
within the mSUGRA scenario, confronting the precise predictions obtained 
after the WMAP results. These data imply, for many of the benchmark points 
retained, a very small difference between the mass of the lightest slepton 
($\tilde{\tau}_1$), the SUSY partner of the $\tau$, 
and the LSP mass ($\Delta M=m_{\tilde{\tau}_1}-m_{\chi}$) since one of 
the preferred mechanism to regulate the amount of DM in the universe is 
the so-called `co-annihilation mechanism'. 

Previous studies have shown that the masses of both smuon and LSP can be
precisely measured using the so-called end-point method [2]. Here
we shall thus concentrating on the measurement of stau mass, which is relevant
as the amount of DM depends critically on it.

Two methods are presented for this purpose. The first is appropriate when 
the stau mass is comparable to the beam energy and the expected cross section 
of the stau production is small. The second works when the stau mass is 
significantly smaller than the beam energy and the stau production cross 
section is large. We also address detectability issues related to different 
collision mode either in head-on or with a half crossing angle of $10$\,mrad.

The end-point method could not applied to the stau analysis as there are
additional missing energies arising from neutrinos in subsequent $\tau$
decays. Furthermore, the final state particle is very soft, 
typically a few GeV for $\Delta M=5$\,GeV taking
benchmark point $D^\prime$ in [4] as a working point.
Another difficulty comes from the fact that the signal cross section is often
many orders of magnitude smaller than that of the Standard Model (SM) 
processes. 

For point $D^\prime$ at center-of-mass energy ($\sqrt{s}$) of
$500$\,GeV, the cross section of the signal process
$e^+e^-\rightarrow \tilde{\tau}_1^+\tilde{\tau}_1^-$ is around $10$\,fb, 
which is to be compared with $10^5-10^6$\,fb of the dominant SM background 
processes $e^+e^-\rightarrow e^+e^-\tau^+\tau^-, e^+e^-\mu^+\mu^-$ and 
$e^+e^-q\overline{q}$. The spectator $e^\pm$ in the background process
is however predominantly peaked in the forward direction. 
Therefore an efficient tagging down to lowest possible
angle is crucial in rejecting these background events. 
Quantitative studies show that the current tagging efficiency of 
the beam monitor calorimeter, LCAL, does not allow for a background free
analysis. Such an analysis may be achieved [1] when the LCAL is fully
efficient in tagging all spectator $e^\pm$ having a transverse momentum 
above $0.8$\,GeV and when additional discriminating variables are used.
One such variable is the scalar sum of transverse momentum with respect to 
the thrust axis in the plane transverse to the beam directions. 
The resulting signal efficiency and background contribution from 
the dominant processes in the head-on collision 
are summarized in table~\ref{tab:effevtd}.
     \begin{table}[ht] 
     \begin{center}
\begin{tabular}{|c|c|c|c|}
\hline 
 Efficiency ($\%$) & $N(\tilde{\tau}_1\rightarrow \tau\chi)$ & 
  $N(ee\rightarrow \tau\tau ee)$ & $N(ee\rightarrow q\overline{q}ee)$ with $q=c,b$ \\
\hline
 $6.3\pm 0.2$ & $316\pm9$ & $1.0\pm 1.0$ & $1.0\pm 1.0$ \\
\hline 
\end{tabular}
     \end{center}
     \caption{\it The efficiency, the signal and dominant background events in
 the head-on case for benchmark point D$^\prime$ at $\sqrt{s}=500$\,GeV.}
\label{tab:effevtd}
     \end{table}

In collisions with a half cross angle of $10$\,mrad, there are two beam holes
for the incoming and outgoing beams. The spectator $e^\pm$ may end up in 
the incoming beam hole resulting additional inefficiency in the veto. 
Studies show that these background events have an unbalanced transverse 
momentum of about $5$\,GeV due to the untagged $e^\pm$ spectator and 
can be efficiently eliminated with by a combined cut on the acoplanarity 
angle and on the angle of the missing transverse momentum vector [1]. 
The price to pay is however a lower signal inefficiency of about $25\%$ 
with respect to the head-on mode.

To extract the $\tilde{\tau}_1$ mass with minimum luminosity, 
the first method consists in measuring the cross section at one energy and 
deduce the mass from the value of $\beta$ since, at the Born level, 
this cross section depends on $\beta^3=(1-4m^2/s)^{3/2}$, where $m$ 
stands for the stau mass. 
Assuming the SM background is negligible and for a given integrated luminosity,
the best accuracy on the stau mass measurement is achieved when the beam 
energy is just above the stau mass threshold [1].
For point D$^\prime$ with $m_{\tilde{\tau}_1}=217$\,GeV, the optimum 
$\sqrt{s}$ is at $\sim 442$\,GeV and the resulting error on the stau mass 
is $\sim 0.5$\,GeV for $500$\,fb$^{-1}$.
The gain in the precision with this choice of optimum beam energy
is appreciable, at $\sqrt{s}=500$\,GeV the error would have been 
$1.2$\,GeV. The same analysis without further optimizing the selection cuts
is also applied to the other relevant benchmark points. The results are
summarized in table~\ref{tab:results}.
     \begin{table}[ht] 
     \begin{center}
\begin{tabular}{|c|c|c|c|c|c|c|c|c|c|}
\hline 
  & \multicolumn{5}{|c|}{method one} & \multicolumn{4}{|c|}{method two}\\
\hline
 Model & A$^\prime$ & C$^\prime$ & D$^\prime$ & G$^\prime$& J$^\prime$ &
 \multicolumn{3}{|c|}{SPS\,1a inspired} & $D^\prime$ \\
\hline $m_{\tilde{\tau}_1}$ (GeV) & $249$ & $167$ & $217$ & $157$ & $312$ & 
 \multicolumn{3}{|c|}{$133$} & $217$ \\
\hline $\Delta M$ (GeV) & $7$ & $9$ & $5$ & $9$ & $3$ & $8$ & $5$ & $3$ & $5$\\
\hline ${\cal L}$ (fb$^{-1}$) & \multicolumn{5}{|c|}{$500$} &
 \multicolumn{3}{|c|}{$200$} & $300$ \\
\hline $\sqrt{s}$ (GeV) & $505$ & $337$ & $442$ & $316$ & $700$ &
 \multicolumn{3}{|c|}{$400$} & $600$ \\
\hline $\sigma$ (fb) & $0.216$ & $0.226$ & $0.279$ & $0.139$ & $1.35$ & 
 \multicolumn{3}{|c|}{$140$} & $50$ \\
\hline $\epsilon$ ($\%$) & $10.4$ & $14.3$ & $5.7$ & $14.4$ & $<1.0$ &
 \multicolumn{3}{|c|}{$18.5$} & $7.6$ \\ 
\hline $\delta m_{\tilde{\tau}_1}$ (GeV) & $0.487$ & $0.165$ & $0.541$ & 
 $0.132$ & $>1.0$ & $0.14$ & $0.22$ & $0.28$ & $0.15$ \\
\hline $\delta (\Omega_{\rm DM}h^2)$ ($\%$) & $3.4$ & $1.8$ & $6.9$ & $1.6$ & 
 $>14$ & $1.7$ & $4.1$ & $6.7$ & $1.9$ \\
\hline 
\end{tabular}
     \end{center}
     \caption{\it Different benchmark points studied in two methods are shown
 together with the $\tilde{\tau}_1$ mass, the mass difference $\Delta M$, 
 the assumed integrated luminosity ${\cal L}$, the chosen center-of-mass 
 energy $\sqrt{s}$, the corresponding signal cross section $\sigma$, 
 the signal efficiency of the selection $\epsilon$, 
 the measured stau mass uncertainty 
 $\delta m_{\tilde{\tau}_1}$ and the resulting precision on DM density
 $\delta (\Omega_{\rm DM}h^2)$.}
\label{tab:results}
     \end{table}

The program Micromegas [5] has been used to compute the relative
uncertainty on the DM density due to the SUSY mass error measurements. 
This program operates without any assumption, in particular it does not 
rely on the mSUGRA scheme. Results listed in table~\ref{tab:results} show, 
as expected, that $\Omega_{\rm DM}h^2$ depends primarily on the precision 
on the stau and LSP masses.
The analysis, optimized for the D$^\prime$ solution, gives satisfactory 
results except for point J$^\prime$ which is almost beyond detectability. 

The second method works at higher beam energies but still below the mass
thresholds of other sparticles, the idea being at such large energies, 
the signal cross section is big enough to collect a large event sample.
The analysis also benefits from using explicitly the polarized beams to
enhance the signal over background ratio.   
The stau mass can then be determined by analyzing the high energy spectrum.
Indeed for the same benchmark point $D^\prime$, if $\sqrt{s}$
could be chosen to be at $600$\,GeV, sufficiently higher than the
stau mass, a more precise stau mass and therefore DM density could be
achieved even with a moderately small integrated luminosity of 
$300$\,pb$^{-1}$.
The same method has also been applied to a SPS\,1a inspired model for 
three different $\Delta M$ values. Again the precision on the stau mass and 
DM density improves as $\Delta M$ increases.

To summarize, our studies have shown that the detection and the mass 
measurement of the tau slepton, potentially important in view of the 
cosmological implications, is challenging in the so-called ``co-annihilation''
scenario. A forward veto to remove the $\gamma\gamma$ background down to 
very small angles is essential to reach an almost background free result, 
adequate to achieve the accuracy implied by the post-WMAP results in 
a model independent analysis. 

In our analysis with method one, we have assumed an ideal detector for
particle detection but with realistic detector acceptance as expected from
a fast simulation program SGV [6], developed and tested at LEP.
Some of the detector capabilities are not yet fully explored, e.g., 
the $dE/dx$ information of the tracking device and possible different
decay lengths (secondary vertex distributions) between the signal and 
the background events.
The analysis in terms of the signal and background separation may still be 
improved using more sophisticated likelihood methods instead of simple cuts.

Nevertheless, the stau analysis in collisions with a crossing angle is 
likely to be more difficult (thought still feasible) than in head-on 
collision, only possible in the TESLA scheme. In a warm machine like
NLC, the same conclusion could be reached provided that 
there is no degradation due to pile-up of several bunches in the forward 
veto (this may require some R$\&$D for a very fast calorimeter). 

\section*{Acknowledgements}
The author is grateful to P.~Bambade, M.~Berggren and F.~Richard for the
fruitful collaboration on the mass threshold method and to U.~Martyn for 
providing material from his study based on analyzing the energy spectrum.


\end{document}